%% file: main.tex
\pdfoutput=1
\documentclass[sigconf, screen, authorversion,nonacm]{acmart}

\AtBeginDocument{%
  \providecommand\BibTeX{{%
    Bib\TeX}}}

\usepackage{svg}
\svgpath{{./assets/}} 
\usepackage{xcolor}
\usepackage[frozencache,cachedir=minted-cache]{minted}
\usepackage{listings}
\usepackage{fontawesome}
\usepackage{tikz, pgfplots}
\pgfplotsset{compat=1.18}
\usepackage{caption}
\usepackage{amsmath}
\usepackage{moresize}

\usetikzlibrary{arrows.meta}
\usetikzlibrary{positioning, calc, matrix, shapes, shadings}

\definecolor{lightlightgray}{gray}{0.95}

\def\BibTeX{{\rm B\kern-.05em{\sc i\kern-.025em b}\kern-.08em
    T\kern-.1667em\lower.7ex\hbox{E}\kern-.125emX}}

\newlength{\mintednumbersep}
\AtBeginDocument{%
  \sbox0{\tiny00}%
  \setlength\mintednumbersep{\parindent}%
  \addtolength\mintednumbersep{-\wd0}%
}

\newcommand{\ischema}[1]{\mintinline[bgcolor=lightlightgray,fontsize=\scriptsize,autogobble=true]{schema-lexer.py:SchemaLexer -x}|#1|}
\newcommand{\oschema}[1]{\mint[bgcolor=lightlightgray,fontsize=\scriptsize]{schema-lexer.py:SchemaLexer -x}|#1|\noindent}
\newenvironment{schema}{%
    \VerbatimEnvironment
    \centering
    \begin{minipage}{1\columnwidth}%
        \begin{minted}[bgcolor=lightlightgray,fontsize=\scriptsize,autogobble=true,linenos,        numbersep=7pt,xleftmargin=15pt, breaklines]{schema-lexer.py:SchemaLexer -x}}
{%
        \end{minted}
    \end{minipage}
    }

\newcommand{\ipycode}[1]{\mintinline[bgcolor=lightlightgray,fontsize=\scriptsize,autogobble=true]{python}|#1|}
\newenvironment{mpycode}[1][]{%
\VerbatimEnvironment
\begin{minted}[bgcolor=lightlightgray,linenos,numbersep=7pt,xleftmargin=15pt, fontsize=\ifblank{#1}{%
    \scriptsize%
  }{%
  #1%
  },autogobble=true]{python}}
{%
\end{minted}}

\newcommand{\attr}[1]{\textsc{#1}}
\newcommand{\rel}[1]{\textsc{#1}}
\newcommand{\node}[1]{{\textsc{#1}}}
\newcommand{\entity}[1]{{\textsc{#1}}}
\newcommand{\code}[1]{\textsc{#1}}

\begin{document}

\include{assets/tikz/colors}

\title[Data2Neo]{Data2Neo - A Tool for Complex Neo4j Data Integration}

\author{Julian Minder}
\authornote{Correspondence to jminder@ethz.ch}
\email{jminder@ethz.ch}
\affiliation{%
  \institution{\textit{Chair of Systems Design,} \textit{ETH Zürich}}
  \city{Zurich}
  \country{Switzerland}
}

\author{Laurence Brandenberger}
\email{lbrandenberger@ethz.ch}
\affiliation{%
  \institution{\textit{Chair of Systems Design,} \textit{ETH Zürich}}
  \city{Zurich}
  \country{Switzerland}
}
\author{Luis Salamanca}
\email{luis.salamanca@sdsc.ethz.ch}
\affiliation{%
  \institution{\textit{Swiss Data Science Center,} \textit{ETH Zürich}}
  \city{Zurich}
  \country{Switzerland}
}\author{Frank Schweitzer}
\email{fschweitzer@ethz.ch}
\affiliation{%
  \institution{\textit{Chair of Systems Design,} \textit{ETH Zürich}}
  \city{Zurich}
  \country{Switzerland}
}

\renewcommand{\shortauthors}{Minder et al.}

\begin{abstract}
This paper introduces Data2Neo, an open-source Python library for converting relational data into knowledge graphs stored in Neo4j databases. 
With extensive customization options and support for continuous online data integration from various data sources, Data2Neo is designed to be user-friendly, efficient, and scalable to large datasets. 
The tool significantly lowers the barrier to entry for creating and using knowledge graphs, making this increasingly popular form of data representation accessible to a wider audience.
The code is available at \href{https://github.com/jkminder/data2neo}{jkminder/data2neo}.

\end{abstract}



\maketitle

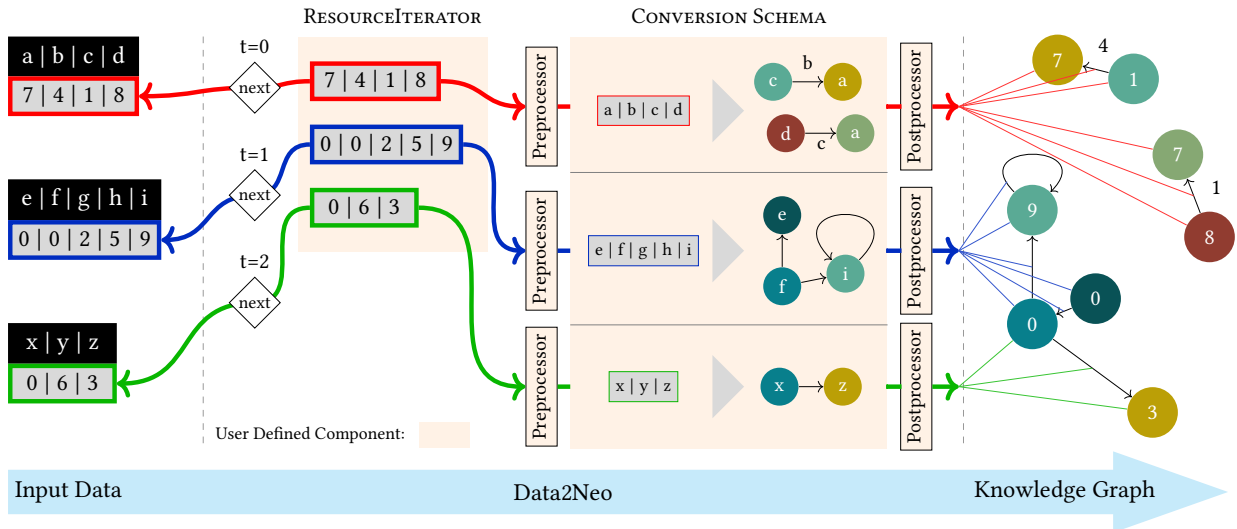
\begin{figure*}[!ht]
    \resizebox{\textwidth}{!}{%
    \input{assets/tikz/data2neo_overview}    
    }%
    
    \caption{A schematic overview of data integration with Data2Neo: We are given three input data sources, each represented by a color, with one row per table. The Figure displays three timesteps ($t=0, t=1, t=2$). The user defines a \textsc{ResourceIterator} that iterates over all the input data and specifies an abstract conversion recipe for each data source within a \textsc{Conversion Schema}. We visualize different node labels by colors. The converted data is pushed to the Knowledge Graph. You can also merge information from different sources – the $f$ and $x$ node from the blue and green data source are merged because they have the same value ($0$) and share the same label (as represented by the color). To customize the integration the user can define pre- and postprocessor functions. These functions support arbitrary Python code. }
    \label{fig:data2neo}
    \Description{A schematic overview of data integration with Data2Neo: We are given three input data sources, each represented by a color, with one row per table. The Figure displays three timesteps ($t=0, t=1, t=2$). The user defines a \textsc{ResourceIterator} that iterates over all the input data as well as an abstract conversion recipes for each data source in a \textsc{Conversion Schema}. We visualize different node labels by colors. The converted data is pushed to the Knowledge Graph. You can also merge information from different sources – the $f$ and $x$ node from the blue and green data source are merged because they have the same value ($0$) and share the samee label (as represented by the color). To customize the integration the user can define pre- and postprocessor functions. These functions support arbitrary Python code.}
\end{figure*}

\section{Introduction}

Relational databases are the most common way of organising data. 
However, they pose severe limitations for data handling because of their static set of tables with a fixed set of columns, each table representing a different entity or concept \cite{ chen2020graph, zhang2011automatic, vyawahare2018hybrid, batra2012comparative,unal2018migration,de2013converting}.
Therefore, graph databases have emerged as a widely accepted substitute for relational databases, with Neo4j as one of the most prominent systems.
In graph databases information is stored in nodes and their relationships.
Any node can store any amount of information and can be connected to any other node through relationships, represented as directed edges between related nodes. 
Additionally, relationships can store any data. 

Such databases are more flexible also because
they can grow organically and support dynamic data. 
Consequently, graph databases have attracted a lot of interest in the scientific community \cite{hogan2021knowledge, ehrlinger2016towards, lin2015learning, paulheim2017knowledge, pujara2013knowledge, wang2017knowledge, wang2014knowledge, noy2019industry}. 
While relational database systems are optimized for \emph{aggregated} data, graph databases are optimized for highly \emph{interlinked} data \cite{miller2013graph}, i.e. they can be visualized in \emph{Knowledge Graphs} (KGs). Depending on the data, KGs have been shown to provide significant advantages in terms of query execution time \cite{relgrapcompsoad, olga_relvgraph, shalini_relvgraph}

KGs are proving to be pivotal in many machine learning applications, including Large Language Models and Retrieval-Augmented Generation (RAG) systems, by enhancing the integration and retrieval of structured knowledge. 
They provide context and background knowledge that is essential for sophisticated applications, such as question answering and advanced automated reasoning across various domains \cite{yasunaga2021qa, wang2023vqa, lewis2020retrieval, he2024g, potts_graphrag_2024, gao2022medml, xie2022unifiedskg, pan2024unifying, chen2024exploring, mai2023opportunities}. 
In addition, they serve as a rich source of semantically structured information for high-quality training data \cite{nickel2015review, agarwal2020knowledge, yasunaga2022deep}. 
Beyond their role in machine learning frameworks, KGs also have intrinsic potential for data analysis in their own right.
They enable the extraction of insights through the relationships they map between different data points, which is critical in domains such as biomedicine \cite{Nicholson2020Constructing, zeng2022toward}, cybersecurity \cite{jia2018practical, liu2022recent}, and financial services \cite{kertkeidkachorn2023finkg}.
The capacity of KGs to seamlessly integrate diverse data types and provide a holistic view makes them invaluable for predictive analytics and decision-making processes.

While certain datasets may opt for graph databases as their storage solution from the outset, there is considerable value in integrating existing relational data into knowledge graphs.  
However, integrating data into KGs might not be straightforward for many interested users, especially when the data integration necessitates employing complicated data transformation pipelines, e.g. when the data needs to be cleaned up or updated in real-time. 
To tackle these challenges, and to lower the entry barrier of data integration, we introduce \textbf{Data2Neo}\footnote{\url{https://github.com/jkminder/data2neo}}, an open-source Python library designed for building data pipelines that convert relational data into knowledge graphs. 
The library provides extensive customization options while remaining easy to use. 
The main contributions of Data2Neo are \emph{i)} abstraction of an easy-to-define YAML-like conversion recipe, \emph{ii)} trivial integration of custom pipeline steps, \emph{iii)} the ability to convert and stream from any data source, and \emph{iv)} optimized parallelized data processing and KG generation.

\section{Existing Solutions}

The official Neo4j documentation mentions several ways to import data into a Neo4j graph database \cite{neo4jimporttut}.

\paragraph{Direct Import \textnormal{(DI)}} Neo4j has native support for importing CSV files, which can be a convenient option for simple conversions. 
This option requires a dump of individual entity tables, converted to nodes, and join-tables, converted to relationships, formatted as CSV files. 
It is the simplest and most efficient way to import data into Neo4j. 
It is possible to add customization through \rel{CYPHER}, Neo4j's query language, but this adds complexity to the process, making it unsuitable for complicated data integration. 
DI is comparable to processing and manipulating data directly via SQL in a relational database.
An extension to this is the \rel{APOC} plugin \cite{neo4japoc} for Neo4j, which adds support for more data sources and various \rel{CYPHER} add-ons, such as conditional \rel{CYPHER} execution. 
While the plugin provides additional functionality, it does not solve the problem of the overhead complexity of \rel{CYPHER} queries for data processing.

\paragraph{Extract-Transform-Load tools \textnormal{(ETL)}} These mostly GUI-based software solutions are designed to handle general data conversion and transformation tasks. 
While there are many ETL options available, the official \emph{Neo4j Developer Guides} \cite{neo4jimporttut} recommend two in particular. 
The first is the native \textbf{Neo4j ETL Tool} \cite{neo4jetltool}  which translates relational data from a JDBC connection to a Neo4j graph. 

Although this tool is well integrated into the Neo4j Desktop application and supports various relational databases, it allows for only limited customization.
For instance, apart from basic type conversions, it lacks the option to add data cleaning steps.
The second one is  \textbf{Apache Hop} \cite{apachehop}, an integration platform for data and metadata orchestration. It scales well to large systems and databases, and provides a visual representation of a data transformation, but its wide range of features can create significant complexity, particularly in research and non-enterprise situations.

\paragraph{Programmatic} The final option is to code the conversion pipeline from scratch. This approach provides full control and the ability to handle arbitrary complexity, but also has significant drawbacks. Developing a data conversion pipeline from scratch is a time-consuming task that quickly becomes overly complicated, especially if the pipeline needs to be dynamically extended or parallelized. Data2Neo targets this context. 

Data2Neo is a lightweight conversion library that bridges the gap between the simplicity of ETL tools and the flexibility of a programmatic approach. It frees the developers from implementing the graph interactions, allowing them to focus on the abstract conversion recipe and data processing using the intuitive options for data conversion and wrangling implemented.

It should be noted that Data2Neo is based on concepts introduced by the \emph{py2neo} Python library \cite{py2neo}, which is deprecated and no longer maintained. The Data2Neo Python library is independent of the \emph{py2neo} library, and integrates some of its functionalities.

\section{Library Overview}

Data2Neo decomposes the data integration process for a user into three distinct parts, shown in Figure~\ref{fig:data2neo}: 
\begin{itemize}
    \item \textsc{ResourceIterator:}  This part iterates over the rows or entities in the input data, which may consist of various data sources, such as multiple tables in a single database or different tables across multiple databases. We call a single row or entity a \textsc{Resource}. It must be assigned a static type such as, for example, the table name in a relational database. \textsc{Resources} with the same type should have the same properties. The \textsc{Resource type} is used to determine how a \textsc{Resource} should be converted. This abstraction provides us with the flexibility to accommodate various data sources.
    \item \textsc{Conversion Schema:} This part specifies the conversion recipe for each \textsc{Resource type}, effectively serving as a blueprint for their integration. Its syntax is an extension to YAML and is intuitive to understand.
    \item \textsc{Converter:} It takes data from the \textsc{ResourceIterator} and integrates it into an existing KG, based on the specifications provided in the \textsc{Conversion Schema}. The \textsc{Converter} automatically parallelizes the processing onto multiple cores, enabling it to scale to large systems. This is the main Python class a user interacts with. 

\end{itemize}

Users define how to iterate over data (\textsc{ResourceIterator}) and how to convert it at an abstract level (\textsc{Conversion Schema}). The \textsc{Converter} then automates the concurrent integration process. Data2Neo greatly facilitates the creation of a KG from multiple data sources and can be a real advantage for users without expertise in graph database technologies.

Furthermore,  Data2Neo excels the most on its customizability.
As sketched in Figure~\ref{fig:data2neo} (middle), users can customize the processing by defining pre- and post-processing functions. A \textsc{preprocessor} adjusts a \textsc{Resource} prior to its processing by the \textsc{Converter}, while a \textsc{postprocessor} alters the output. These functions, referred to as "wrappers", offer limitless customization possibilities and can incorporate any Python code. For example, they can be used to filter and clean data, enrich data with additional information, verify and moderate data, and more.

\subsection{Sample Task}

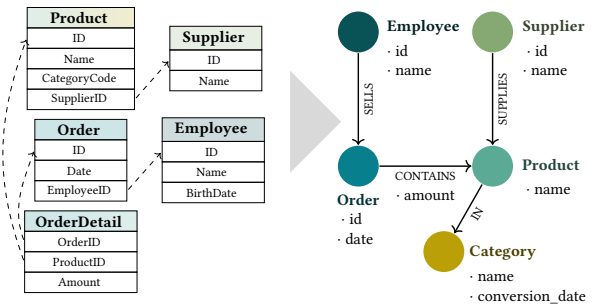
\begin{figure}
    \setlength{\abovecaptionskip}{0pt}
    \centering
    \resizebox{1\linewidth}{!}{
        \input{assets/tikz/example_conversion}    
    }
    \caption{Example Conversion Task: A simple product sales system is converted into a knowledge graph.}    \Description{Example Conversion Task: A simple product sales system is converted into a knowledge graph.}
    \label{fig:example}
\end{figure}

We present a sample task to demonstrate the library's API. We examine a simple retail sales system consisting of five relational database tables:  \entity{Product}, \entity{Order}, \entity{Supplier}, \entity{Employee} and \entity{OrderDetail}. 
When using this information for downstream analysis tasks, the synergies and relationships between these different actors are essential. For this reason, a knowledge graph, as the one shown in Figure \ref{fig:example}, is the representation that enables richer and more complex exploration. 

We transform each entity from the \entity{Supplier}, \entity{Order}, and \entity{Employee} tables into individual nodes. Additionally, within the \entity{Order} table, each entity contains a unique ID corresponding to the employee responsible for creating the {order}. 
This foreign key is transformed into a relationship linking the \node{Employee} and \node{Order} nodes. 
The \entity{Product} entities are divided into two distinct nodes: the \node{Product} node itself and a \node{Category} node linked to the \node{Product} node via an \rel{IN} relationship. 
To ensure data integrity, we merge \node{Category} nodes based on the \attr{CategoryCode} attribute, consolidating them into a single node for each unique category code.  
Lastly, the \entity{OrderDetail} entities are converted into a relationship between the \node{Order} and \node{Product} nodes, detailing the quantity of product in each order. 
To showcase the library's customization capabilities, we make two enhancements to the \node{Category} node. 
First, the \attr{CategoryCode} is translated to its string name. Second, we parse the parent Category, corresponding to the first \attr{CategoryCode} digit, and add its translated string name as a label to the \node{Category} node. Even though these examples may seem simple, they are impossible, or require much more involved coding, when using existing methods because they require external data input and custom logic.

\subsection{Workflow}

In Listing \ref{lst:maininterface} we present the standard steps required in Data2Neo to carry out the population of the knowledge graph. The \textsc{Converter} is the main Python object that the user interacts with, and needs to be initialized with the following three elements: an \textsc{ResourceIterator} declared according to the data sources, a \textsc{Conversion Schema} that details the mapping from entities to nodes and relationships, and the graph credentials. Once the \textsc{Converter} is instantiated, the user invokes it to start the conversion, which processes the data in two steps. First, it iterates over all \textsc{Resources} to create the nodes in the Neo4j graph, and then it creates the relationships in a second iteration. By default, the \textsc{Converter} uses multiple processes to speed up the conversion process by dividing the \textsc{Resources} into batches and distributing them among the available processes. The transfer of data to the graph is always serialized to ensure correctness. 

In the example presented in Listing \ref{lst:maininterface} we presume that the data is stored in a SQLite database. For SQLite databases and pandas \cite{pandas} DataFrames, Data2Neo provides a \textsc{ResourceIterator} natively; for other databases, users can customize their own \textsc{Resource} and \textsc{ResourceIterator}. The library offers the opportunity to connect multiple iterators together, regardless of their data source.

\begin{listing}

\begin{mpycode}[]
from tqdm import tqdm
from data2neo import Converter
from data2neo.relational_modules.sqlite
    import SQLiteIterator

# Connect to the SQLite database and create iterator
iterator = SQLiteIterator(sqlite3.connect(...))

# define schema
schema = ... # Conversion Schema as string

# define graph credentials
uri, auth  = ... # (host, (user, password))

# setup Converter
converter = Converter(schema, iterator, uri, auth)

# run pipeline
converter(progress_bar = tqdm)
\end{mpycode}
\caption{The Python interface of Data2Neo. We connect to a sqlite3 database, for which Data2Neo provides a \textsc{ResourceIterator} out of the box.}
\label{lst:maininterface}
\end{listing}

\subsection{Schema Syntax}
\begin{listing}
\caption{\textsc{Conversion Schema} for sample task.}
\label{schema:full}
\begin{schema}
ENTITY("Product"):
  NODE("Product") productnode:
    + name = Product.Name

  ParseParentCategory(NODE("Category", CodeToCategory(Product.ParentCategory))) categorynode:
    + name = CodeToCategory(Product.CategoryCode)
    - conversion_date = Product.ConversionDate
  
  RELATIONSHIP(productnode, "IN", categorynode):
  RELATIONSHIP(MATCH("Supplier", id=Product.SupplierID), "SUPPLIES", productnode):

ENTITY("Order"):
    NODE("Order") ordernode:
      + id = Order.ID
      - date = Order.Date

    RELATIONSHIP(MATCH("Employee", id=ordernode.EmployeeID), "SELLS", ordernode):

ENTITY("Supplier"):
    NODE("Supplier"):
      + id = Supplier.ID
      - name = Supplier.Name

ENTITY("Employee"):
    NODE("Employee"):
      + id = Employee.ID
      - name = Employee.Name

ENTITY("OrderDetail"):
    RELATIONSHIP(MATCH("Order", id=OrderDetail.OrderID), "CONTAINS", MATCH("Product", id=OrderDetail.ProductID):
      - amount = OrderDetail.Amount
\end{schema}
\end{listing}

The \textsc{Conversion Schema} is an user-defined recipe that allows to declare how relational entities are converted to graph elements (nodes, relationships and their attributes, which correspond to key-value pairs stored on a node or relationship). Each \code{Resource} passed to the \code{Converter} has an associated type, and conversion rules are specified for each type. The schema allows for one-to-one, one-to-many and many-to-one conversions of relational entities to graph elements. 

The full \textsc{Conversion Schema} for the sample task is shown in Listing \ref{schema:full}, with which we illustrate the main features provided by Data2Neo, required to build and populate the KG.

The schema follows a YAML-like syntax. There are four main keywords: \ischema{ENTITY}, \ischema{NODE}, \ischema{RELATIONSHIP} and \ischema{MATCH}. Each keyword is followed by parentheses containing keyword-specific definitions. At the first indent level, we define the \textsc{Resource type} using the \ischema{ENTITY} keyword, and the specific \textsc{Resource type} in quotes. Following YAML syntax the line must end with a colon.
\oschema{ENTITY("Product"):   conversion definitions...}
Now, different attributes of the \textsc{Resource} can be specified using the entity name as prefix, e.g.,~\ischema{Product.Name}. These attributes from the \ischema{ENTITY} can be used to define nodes and relationships.

\paragraph{Nodes} To generate nodes from input data, we use the \ischema{NODE} keyword. In parentheses, we specify the comma-separated labels. In Neo4j a label is a type associated with a node. Each node can have multiple labels. We can also use an attribute of the \textsc{Resource} as a label:
\oschema{NODE("Label", "Other Label", Product.Name)}

After the parentheses, we can define an optional identifier with which we can reference the node when defining Relationships. In lines 2-3 of Listing \ref{schema:full}, \ischema{productnode} serves as the identifier. See line 9 for its reference when defining a relationship; more details on defining relationships are provided later.

To set attributes of a node, we follow standard YAML syntax. The attribute name can be different from that of the \textsc{Resource} attribute, as we can observe in line 3 where \ischema{name} is used for attribute \ischema{Name} of \ischema{Product}. Also in line 3 we can observe how \ischema{-} has been replaced by a \ischema{+} to specify it is the primary attribute of the node. A primary attribute is a unique, non-null identifier assigned to a node. Primary attributes as well as the first specified label are used to merge nodes, i.e.~if the primary attribute of two nodes is the same and they have an identical first label, they are merged. When merging elements, Data2Neo will update all the attributes and labels of the existing element with the new information. This is useful to push information from different \textsc{Resources} to the same node, allowing for complex many-to-one conversions. However, it is important to bear in mind that primary attributes can slow down the conversion process as it requires additional computation for matching. 

\paragraph{Relationships} 
The \ischema{RELATIONSHIP} keyword requires three arguments: the source node, the type of the relationship, and the target node: 

\oschema{RELATIONSHIP(-source-, "TYPE", -target-):}

The reference to the \ischema{-source-} and \ischema{-target-} can be done either through node identifiers or with a \ischema{MATCH} clause, which allows matching any set of nodes in the graph. If the source or the target has multiple nodes, a relationship is created for each combination of source and target. The \ischema{MATCH} clause expects one or more comma-separated labels, followed by zero or more attribute equalities. For example, in line 10 we match a \node{Supplier} node in the graph whose \attr{id}  attribute matches the value of \ischema{Product.SupplierID}, and create a relationship of type \ischema{SUPPLIES} between these nodes and the node with identifier \ischema{productnode}. Attributes are defined in the same way as they are for nodes, and similarly, primary attributes can be used to merge relationships.

\subsection{Customization with Wrappers}

Wrappers are a fundamental component of the Data2Neo library, enabling users to enhance its functions and tailor it to their individual requirements. 

After the user has defined the three elements required -- a \textsc{ResourceIterator}, a \textsc{Conversion Schema}, and the  \textsc{Converter} -- the process of building and populating the KG can be started. The  \textsc{Converter} compiles the blueprints, outlined in the \textsc{Conversion Schema} into a conversion pipeline. This pipeline processes each \textsc{Resource} in two main steps.  First, a subgraph, a graph formed by just the nodes and/or relationships from one single \textsc{Resource}, is built. Then, the attributes of the nodes and relationships in the subgraph are filled in one at a time. Pre- and postprocessors, built as Python functions, can be used in each of these steps. They allow to customize either the input or output data, acting before (preprocessor) or after (postprocessor) a \textsc{Resource} is processed in each step. A subgraph postprocessor acts on the nodes and relationships of a subgraph, while an attribute postprocessor manipulates the attributes, i.e., key-value pairs, of nodes and relationships. A node label or relationship type is also treated as an attribute, except its key is empty. Both a subgraph preprocessor and an attribute preprocessor operate on the input \textsc{Resource}.

This system offers considerable flexibility. Although modifications can often be made in more than one way, either before or after data conversion, this two points intervention capability allows for a high degree of customization, accommodating any user's specific needs.
It is important to remark that defining a wrapper is as easy as registering a Python function with Data2Neo through a decorator. Afterward, the wrapper's name can be referenced in the \textsc{Conversion Schema}.

In our example scenario, we convert the category code of the \node{Category} node to a name. We define the function \ipycode{CodeToCategory} and register it as \emph{attribute} postprocessor. We also modify the \textsc{Resource} to include the parent category using the \ipycode{ParseParentCategory} \emph{subgraph} preprocessor. We can reuse the \ipycode{CodeToCategory} wrapper to convert this to its string version. See Listing \ref{lst:wrappers} for the definition of the wrappers and lines 5 and 6 in Listing \ref{schema:full} for their use in the schema. Besides, let's note that we have the freedom to use any other Python tools, including state-full wrappers. For further details, please consult the \href{https://data2neo.jkminder.ch}{documentation}. 

\begin{listing}
\begin{mpycode}[\scriptsize]
from data2neo import register_attribute_postprocessor
from data2neo import register_subgraph_preprocessor
from data2neo import Attribute

@register_attribute_postprocessor
def CodeToCategory(attribute):
    # an attribute is defined by its key and a value
    code = attribute.value
    conversion = {
        1: "Clothing",
        2: "Home appliances",
        101: "T-Shirts",
        102: "Pants",
        ...
    } # You could also query an API here
    return Attribute(attribute.key, conversion[code])

@register_subgraph_preprocessor
def ParseParentCategory(resource):
    resource["ParentCategory"] = int(str(resource["CategoryCode"])[0])
    return resource
\end{mpycode}
\caption{Definitions of wrappers for the sample task. The attribute postprocessor \ipycode{CodeToCategory} converts the category code to its string corespondent. The \ipycode{AddConversionDate} subgraph preprocessor adds the parent category, which is the first digit of the \textsc{CategoryCode}. as an attribute to the \textsc{Resource}. This can then be accessed in the schema as any other \textsc{Resource} attribute.}
\label{lst:wrappers}
\end{listing}

\section{Performance Analysis}
We compare the runtime performance of the Data2Neo library to \emph{Direct Import} (DI) using \rel{CYPHER} and CSV files. To analyse the runtime of the two methods, we present a new conversion task.

\subsection{Task}

The benchmark uses open-source Github data from \cite{githubdataset} as the data source. For each repository, the data is stored in a relational database with two tables: \textsc{commits} and \textsc{edits}. We create nodes for each commit, file, and author. 
Edits are represented as relationships from the commit to the file with the relationship type given by the \attr{edit\_type} attribute of the edit column in all upper letters. If an edit renames a file, we also create a \rel{RENAMED\_TO} relation to the new file. 

The task requires conditionals, dynamic type setting and attribute modification. In pure \rel{CYPHER} this is not possible, so we need to complement \rel{CYPHER} queries with extra functionalities from the \rel{APOC} plugin. The use of \rel{APOC} requires further expertise. The task has moderate difficulty and is suitable for testing the performance of Data2Neo and DI.

\paragraph{Methodology} We conduct our evaluation on a machine equipped with an Intel Xeon (Skylake) CPU, 64 cores, and 256 GB of RAM. We run a local instance of Neo4j version 5.18.1, using the default configuration with the exception of \texttt{db.import.csv.buffer\_size}, which was set to 1GB to accommodate large input CSV files. For data import with Data2Neo, we utilized 32 workers and a batch size of 20,000. Similarly, for DI, we used transactions of 20,000 rows. Each experiments is repeated 5 times. The input data is stored in an SQLite database. For the DI, we need to first export the database to CSV to be readable via \rel{CYPHER}. Data2Neo can directly access the SQLite database. The benchmark scripts, including the \textsc{Conversion Schema} and the \rel{CYPHER} commands, are available at \href{https://github.com/jkminder/data2neo-performance-evaluation}{jkminder/data2neo-performance-evaluation}.

\subsection{Analysis}
We compare the performance of Data2Neo to that of DI across three different repository sizes.
To run DI, one must install the \rel{APOC} plugin.
Further, one needs to adjust neo4j configurations to handle large CSV inputs.
The results, presented in Table \ref{tab:runtime}, demonstrate that DI is consistently faster due to its utilization of Neo4j's internal structures and performance optimizations. However, Data2Neo, as a pure Python library, still efficiently processes large datasets, handling up to 34 million rows in under 2 hours. Notably, Data2Neo offers a much simpler and more intuitive approach compared to executing this task by first exporting the database and then leveraging \rel{CYPHER} and \rel{APOC}. The \rel{CYPHER} command alone is about 2,500 characters long, whereas the Data2Neo conversion scheme is only about 1,000 characters long, and is much more intuitive.

In summary, while it may not match DI in speed, Data2Neo provides greater flexibility, adaptability, and simplicity, making it a valuable tool for developers and data scientists.
\begin{table}[!ht]
\caption{Conversion runtime comparison for three repositories on the described task. The table includes the number of input rows, as well as the number of nodes and relationships in the created graph. The runtime is displayed in minutes, representing the mean and standard deviation across five experiment repetitions.}
\label{tab:runtime}
\centering
\begin{footnotesize}
\begin{tabular}{l|l|l||l|l}
\multicolumn{1}{l}{Repository} & \multicolumn{1}{l}{ \tiny Input} & \multicolumn{1}{l}{ \tiny Output} &  \multicolumn{2}{l}{Runtime [min]}     \\
\hline
{\footnotesize Name } & \multicolumn{1}{l|}{\footnotesize Rows} & \multicolumn{1}{l||}{\footnotesize Nodes~/~Relations} &  \multicolumn{1}{l|}{\footnotesize DI} & {\footnotesize Data2Neo}\\
\hline

{\scriptsize ansible/ansible}    &  {\scriptsize\phantom{1}6\,917\,313} & {\scriptsize\phantom{1}86\,755~/~\phantom{1}6\,968\,631 }    & \phantom{1}8.7 {\tiny $\pm 0.1$} &   \phantom{1}23.5   {\tiny $\pm 0.4$}     \\
{\scriptsize tensorflow/tensorflow}    &  {\scriptsize 22\,707\,206} & {\scriptsize 135\,234~/~22\,808\,448 }    & 28.1 {\tiny$\pm 2.0$} &   \phantom{1}75.3 {\tiny $\pm 1.7$}       \\
{\scriptsize openshift/origin}      &  {\scriptsize 34\,846\,906 }          & {\scriptsize 125\,515~/~34\,889\,186 }       &    39.7  {\tiny$\pm 0.7$}                      &    112.9  {\tiny$\pm 2.9$}      \\
\end{tabular}
\end{footnotesize}
\end{table}

\section{Conclusion}

We present Data2Neo, an open-source Python library that enables the seamless creation of conversion pipelines from relational data to Neo4j graph databases. Data2Neo simplifies the process by minimizing boilerplate code while still allowing for complex conversion scenarios and enabling continuous online integration of data from any source. Its flexibility makes it an ideal solution for users who have a dynamic research environment and need to regularly update their graph structure and processing functions. The library separates the conversion process into a clear and easily modifiable \textsc{Conversion Schema} and the processing code itself, making it easy to manage and update conversion pipelines. It combines the simplicity of ETL tools like the \emph{Neo4j ETL Tool} with the limitless customization options of a programmatic approach in Python.

In addition to its ease of use, Data2Neo has proven to be highly efficient and scalable, even for large datasets. Despite its Python backend, we have shown that its runtime overhead is manageable, even when compared to  native Neo4j import functions, and it can process millions of records per hour with automatic parallelization. Overall, Data2Neo is a valuable tool for developers and data scientists looking to convert relational data to knowledge graphs.

\bibliographystyle{ACM-Reference-Format}
\bibliography{literature}

\end{document}

%% file: assets/tikz/colors.tex


\definecolor{a}{HTML}{095256}
\definecolor{b}{HTML}{087F8C}
\definecolor{c}{HTML}{5AAA95}
\definecolor{d}{HTML}{86A873}
\definecolor{e}{HTML}{BB9F06}
\definecolor{f}{HTML}{913C30}
\definecolor{darkblue}{HTML}{002CC0}
\definecolor{darkgreen}{HTML}{08B600}

%% file: assets/tikz/data2neo_overview.tex
\newcommand{\dista}{0.7cm}
\newcommand{\distb}{0.7cm}

\newcommand{\internodedist}{2.2cm}
\newcommand{\intertabledist}{0.5cm}
\newcommand{\rellabel}[2]{\textbf{\scriptsize #1} \\ {\scriptsize #2}}
\newcommand{\kgnode}[6]{
    \ifstrequal{#2}{below}{
        \node [fill=#6, label={[align=left,label distance=-3mm]#2:\textcolor{#6!40!black}{\textbf{#1}} \\ \scriptsize #3 },#4] (#1) #5 {};
    }{
        \node [fill=#6, label={[align=left]#2:\textcolor{#6!40!black}{\textbf{#1}} \\ \scriptsize #3 },#4] (#1) #5 {};
    }
}
\newcommand{\resource}[7]{
    \node [draw=#7, rectangle, minimum width = 1cm*#2, minimum height = 0.2cm*#2, fill=#3,#4] (#1) #5 {#6}
}
\newcommand{\rowHeight}{0.5cm}
\newcommand{\rowWidth}{2cm}
\newcommand{\tableshift}{0.3cm}

\begin{tikzpicture}
    \begin{scope}
        \node [draw, line width=2pt, font=\Large, text=white, fill=black!100, minimum width=\rowWidth, minimum height=\rowHeight] (a0) at (0,0) {a $\lvert$ b $\lvert$ c $\lvert$ d};
        \node [draw, draw=red, fill=gray!30, line width=2pt, font=\Large, minimum width=\rowWidth, minimum height=\rowHeight,below=0cm of a0.south] (a1) {7 $\lvert$ 4 $\lvert$ 1 $\lvert$ 8};
       
        \node [draw, line width=2pt, font=\Large,  text=white, fill=black!100, minimum width=\rowWidth+10, minimum height=\rowHeight, below=1cm of a1.south west, anchor=north west] (b0) {e $\lvert$ f $\lvert$ g $\lvert$ h $\lvert$ i};
        \node [draw=darkblue, font=\Large, line width=2pt, text=black, fill=gray!30, minimum width=\rowWidth+10, minimum height=\rowHeight, below=0cm of b0.south] (b1) {0 $\lvert$ 0 $\lvert$ 2 $\lvert$ 5 $\lvert$ 9};

        \node [draw, line width=2pt, font=\Large,  text=white, fill=black!100, minimum width=\rowWidth-10, minimum height=\rowHeight, below=1cm of b1.south west, anchor=north west] (c0) {x $\lvert$ y $\lvert$ z};
        \node [draw=darkgreen, font=\Large, line width=2pt, text=black, fill=gray!30, minimum width=\rowWidth-10, minimum height=\rowHeight, below=0cm of c0.south] (c1) {0 $\lvert$ 6 $\lvert$ 3};
    \end{scope}
    \coordinate[right=1cm of a0.north east] (l1a) ;
    \coordinate[below right=7cm and 1cm of a0.north east] (l1b) ;


    \begin{scope}
    \node [fill=orange!10, rectangle, minimum width=3cm, minimum height=3.4cm,  below right=0.0cm and 2.5cm of a0.north east] (ResourceIt) {};
    \node[font=\large, above=0.1cm of ResourceIt.north]{\textsc{ResourceIterator}};
    
    \node [draw, draw=red, fill=gray!30, line width=2pt, font=\Large, minimum width=\rowWidth, minimum height=\rowHeight, anchor=west] at ($(ResourceIt.west)-(-0.2,-1)$) (ra1) {7 $\lvert$ 4 $\lvert$ 1 $\lvert$ 8};

     \node [draw=darkblue, font=\Large, line width=2pt, text=black, fill=gray!30, minimum width=\rowWidth+10, minimum height=\rowHeight, anchor=west] at ($(ResourceIt.west)-(-0.2,0)$) (rb1) {0 $\lvert$ 0 $\lvert$ 2 $\lvert$ 5 $\lvert$ 9};
     
    \node [draw=darkgreen, font=\Large, line width=2pt, text=black, fill=gray!30, minimum width=\rowWidth-10, minimum height=\rowHeight, anchor=west] at ($(ResourceIt.west)-(-0.2,1)$) (rc1) {0 $\lvert$ 6 $\lvert$ 3};
      
\end{scope}
\tikzstyle{graphnode} = [font=\normalsize, text=white, minimum size=0.6cm, circle]


\newcommand{\convertercolor}{black!60}

\node [fill=orange!10, rectangle, minimum width=50mm, minimum height=65mm, below right=0cm and 1.3cm of ResourceIt.north east, anchor=north west] (Schema) {};


\draw[dashed, black!40] ($(Schema.north west) - (5.8,0)$) -- ($(Schema.south west) - (5.8,0)$);    

\draw[dashed, black!40] ($(Schema.north east) + (1.2,0)$) -- ($(Schema.south east) + (1.2,0)$);   

\coordinate (mida) at ($(ra1.east)!0.43!(a1.west)$);
\draw[-, red, line width=2pt] ($(ra1.west)-(0,0)$) to[in=0, out=-180, looseness=1.3] (mida);
\draw[->, red, line width=2pt] (mida) to [in=0, out=-180, looseness=1.3] (a1.east);
\node[diamond, draw, fill=white, font=\small, inner sep=1pt] at (mida) {next};
\node[font=\large, inner sep=1pt, above=0.5cm of mida] {t=0};

\coordinate (midb) at ($(mida)-(0,1.7)$);
\draw[-, darkblue, line width=2pt] ($(rb1.west)-(0,0)$) to[in=0, out=-180, looseness=1.3] (midb);
\draw[->, darkblue, line width=2pt] (midb) to [in=0, out=-180, looseness=1.3] (b1.east);
\node[diamond, draw, fill=white, font=\small, inner sep=1pt] at (midb) {next};
\node[font=\large, inner sep=1pt, above=0.5cm of midb] {t=1};

\coordinate (midc) at ($(midb)-(0,1.7)$);
\draw[-, darkgreen, line width=2pt] ($(rc1.west)-(0,0)$) to[in=0, out=-180, looseness=1.3] (midc);
\draw[->, darkgreen, line width=2pt] (midc) to [in=0, out=-180, looseness=1.3] (c1.east);
\node[diamond, draw, fill=white, font=\small, inner sep=1pt] at (midc) {next};
\node[font=\large, inner sep=1pt, above=0.5cm of midc] {t=2};

\node[font=\large, above=0.1cm of Schema.north]{\textsc{Conversion Schema}};

\tikzstyle{bigarrow} = [single arrow, minimum width=1cm, minimum height=0.5cm, single arrow head extend=0.2cm, fill=gray!30]

\node [draw, rectangle, minimum width = 1cm, minimum height = 0.2,rotate=90, fill=orange!10, anchor=south] at ($(Schema.north west)!0.17!(Schema.south west)-(0.2,0)$)(PreA) {Preprocessor};

\node [draw, rectangle, minimum width = 1cm, minimum height = 0.2,rotate=90, fill=orange!10, anchor=south] at ($(Schema.north west)!0.52!(Schema.south west)-(0.2,0)$)(PreB) {Preprocessor};

\node [draw, rectangle, minimum width = 1cm, minimum height = 0.2,rotate=90, fill=orange!10, anchor=south] at ($(Schema.north west)!0.85!(Schema.south west)-(0.2,0)$) (PreC) {Preprocessor};

\node [draw, rectangle, minimum width = 1cm, minimum height = 0.2,rotate=90, fill=orange!10, anchor=north] at ($(Schema.north east)!0.17!(Schema.south east)+(0.2,0)$) (PostA) {Postprocessor};

\node [draw, rectangle, minimum width = 1cm, minimum height = 0.2,rotate=90, fill=orange!10, anchor=north] at ($(Schema.north east)!0.52!(Schema.south east)+(0.2,0)$) (PostB) {Postprocessor};

\node [draw, rectangle, minimum width = 1cm, minimum height = 0.2,rotate=90, fill=orange!10, anchor=north] at ($(Schema.north east)!0.85!(Schema.south east)+(0.2,0)$) (PostC) {Postprocessor};

\resource{ResourceA}{0.8}{gray!30}{font=\small}{at ($(Schema.north)!0.18!(Schema.south)-(1.35,0)$)}{a $\lvert$ b $\lvert$ c $\lvert$ d}{red};
\node [graphnode, fill=e, above right=0mm and \distb*2+8mm of ResourceA] (ResourceCNodeA) {a};
\node [graphnode, fill=f, below right=-1mm and \distb+6mm of ResourceA] (ResourceCNodeB) {d};
\node [graphnode, fill=d, below right=-1mm and \distb*2+10mm of ResourceA] (ResourceCNodeC) {a};
\node [graphnode, minimum size=0.6cm, circle, fill=c, above right=0mm and \distb-2mm+6mm of ResourceA] (ResourceCNodeD) {c};

\node[bigarrow] at ($(Schema.north)!0.18!(Schema.south)+(-0.1,0)$) {};
\draw[->] (ResourceCNodeD) to (ResourceCNodeA);
\draw[->] (ResourceCNodeB) to (ResourceCNodeC);
\node [inner sep=2pt, anchor=center] at ($(ResourceCNodeD.north)!0.5!(ResourceCNodeA.north)$) {b};
\node [inner sep=2pt, anchor=center] at ($(ResourceCNodeB.south)!0.4!(ResourceCNodeC.east)$) {c};

\draw[-, black!50] ($(Schema.north west)!0.33!(Schema.south west)$) to ($(Schema.north east)!0.33!(Schema.south east)$);

\resource{ResourceB}{0.8}{gray!30}{font=\small}{at ($(Schema.north)!0.52!(Schema.south)-(1.35,0)$)}{e $\lvert$ f $\lvert$ g $\lvert$ h $\lvert$ i}{darkblue};

\node[bigarrow] at ($(Schema.north)!0.52!(Schema.south)+(-0.1,0)$) {};

\node [graphnode, fill=a, above right=1mm and \distb*2-3mm of ResourceB] (ResourceBNodeA) {e};
\node [graphnode, fill=b, below right=1mm and \distb*2-3mm of ResourceB] (ResourceBNodeB) {f};
\node [graphnode, fill=c, below right=-1mm and \distb*3 of ResourceB] (ResourceBNodeC) {i};
\coordinate (targetloop) at ($(ResourceBNodeC.north) + (-0.25cm, 0.2cm)$);

\draw[->] (ResourceBNodeB) to (ResourceBNodeA);
\draw[->] (ResourceBNodeB) to (ResourceBNodeC);
\path [->, out=45, in=135, looseness=8] (ResourceBNodeC) edge (ResourceBNodeC);

\draw[-, black!50] ($(Schema.north west)!0.70!(Schema.south west)$) to ($(Schema.north east)!0.70!(Schema.south east)$);

\resource{ResourceC}{0.8}{gray!30}{font=\small}{at ($(Schema.north)!0.85!(Schema.south)-(1.35,0)$)}{x $\lvert$ y $\lvert$ z }{darkgreen};

\node[bigarrow] at ($(Schema.north)!0.85!(Schema.south)+(-0.1,0)$) {};

\node [graphnode, fill=b, above right=0mm and \distb*2+2mm of ResourceC.east, anchor=center] (ResourceANodeA) {x};
\node [graphnode, fill=e, above right=0mm and \distb*2 + 12mm of ResourceC.east, anchor=center] (ResourceANodeB) {z};
\draw[->] (ResourceANodeA) to (ResourceANodeB);

\node [anchor=north, below=0.1cm of Schema.north] (Title) {};

\draw[->, red, line width=2pt] ($(ra1.east)-(0,0)$) to[in=180, out=0, looseness=1.3] (PreA.north);
\draw[-, red, line width=2pt] (PreA.south) to[in=180, out=0, looseness=1.3] ($(Schema.north west)!0.17!(Schema.south west)$);
\draw[-, red, line width=2pt] (PostA.north) to[in=180, out=0, looseness=1.3] ($(Schema.north east)!0.17!(Schema.south east)$);

\draw[->, darkblue, line width=2pt] ($(rb1.east)-(0,0)$) to[in=180, out=0, looseness=1.3] (PreB.north);
\draw[-, darkblue, line width=2pt] (PreB.south) to[in=180, out=0, looseness=1.3] ($(Schema.north west)!0.52!(Schema.south west)$);
\draw[-, darkblue, line width=2pt] (PostB.north) to[in=180, out=0, looseness=1.3] ($(Schema.north east)!0.52!(Schema.south east)$);

\draw[->, darkgreen, line width=2pt] ($(rc1.east)-(0,0)$) to[in=180, out=0, looseness=1.3] (PreC.north);
\draw[-, darkgreen, line width=2pt] (PreC.south) to[in=180, out=0, looseness=1.3] ($(Schema.north west)!0.85!(Schema.south west)$);
\draw[-, darkgreen, line width=2pt] (PostC.north) to[in=180, out=0, looseness=1.3] ($(Schema.north east)!0.85!(Schema.south east)$);


\tikzstyle{graphnode} = [font=\large, text=white, minimum size=0.8cm, circle]

\coordinate (GraphCenter) at ($(Schema.north east)!0.5!(Schema.south east) + (2,-0.2)$);
\node [font=\small, graphnode,  fill=e, above right=28mm and \distb*2-10mm of GraphCenter] (NodeEA) {7};
\node [font=\small,graphnode, fill=f, above right=0mm and \distb*4 of GraphCenter] (NodeFA) {8};
\node [font=\small,graphnode,  fill=d, above right=13mm and \distb*3+2mm of GraphCenter] (NodeDA) {7};
\node [font=\small,graphnode,  fill=c, above right=25mm and \distb*2+2mm of GraphCenter] (NodeCA) {1};
    
\draw[->, black] (NodeCA) to (NodeEA);
\draw[->, black] (NodeFA) to (NodeDA);
\node [inner sep=2pt, font=\large, anchor=center] at ($(NodeCA.north)!0.4!(NodeEA.north)$) {4};
\node [inner sep=2pt, font=\large,  anchor=center] at ($(NodeFA.east)!0.6!(NodeDA.east)$) {1};

\node [graphnode, fill=a, below right=4mm and 10mm of GraphCenter] (NodeAB) {0};
\node [graphnode, fill=b, below right=8mm and 0mm of GraphCenter] (NodeBB) {0};
\node [graphnode,fill=c, below right=-10mm and 0mm of GraphCenter] (NodeCC) {9};

\draw [->, draw=black] (NodeBB) to (NodeCC);
\path [->, out=135, in=45, looseness=5,draw=black] (NodeCC) edge (NodeCC);
\draw [->, draw=black] (NodeAB) to (NodeBB);

\node [graphnode, fill=e, below right=22mm and 19mm of GraphCenter] (NodeEC) {3};
\draw [->] (NodeBB) to (NodeEC);

\coordinate (GenerationSourceA) at ($(PostA.south)+(0.42,0)$);
\newcommand{\opp}{!80}
\draw[->, red, line width=2pt] ($(PostA.south)-(0,0)$) to[in=180, out=0, looseness=1.3] (GenerationSourceA);
\draw[-,  draw=red\opp] (GenerationSourceA) to (NodeEA);
\draw[-,  draw=red\opp] (GenerationSourceA) to (NodeFA);
\draw[-,  draw=red\opp] (GenerationSourceA) to (NodeDA);
\draw[-,  draw=red\opp] (GenerationSourceA) to (NodeCA);
\draw[-,  draw=red\opp] (GenerationSourceA) to ($(NodeCA)!0.5!(NodeEA)$);
\draw[-,  draw=red\opp] (GenerationSourceA) to ($(NodeFA)!0.5!(NodeDA)$);

\coordinate (GenerationSourceB) at ($(PostB.south)+(0.42,0)$);
\draw[->, darkblue, line width=2pt] ($(PostB.south)-(0,0)$) to[in=180, out=0, looseness=1.3] (GenerationSourceB);
\draw[-, draw=darkblue\opp] (GenerationSourceB) to (NodeAB);
\draw[-,  draw=darkblue\opp] (GenerationSourceB) to (NodeBB);
\draw[-,  draw=darkblue\opp] (GenerationSourceB) to (NodeCC);
\draw[-,  draw=darkblue\opp] (GenerationSourceB) to ($(NodeBB)!0.5!(NodeCC)$);
\draw[-,  draw=darkblue\opp] (GenerationSourceB) to ($(NodeAB)!0.5!(NodeBB)$);
\draw[-,  draw=darkblue\opp] (GenerationSourceB) to ($(NodeCC)-(0.4,-0.45)$);

\coordinate (GenerationSourceC) at ($(PostC.south)+(0.42,0)$);
\draw[->, darkgreen, line width=2pt] ($(PostC.south)-(0,0)$) to[in=180, out=0, looseness=1.3] (GenerationSourceC);
\draw[-, draw=darkgreen\opp] (GenerationSourceC) to (NodeBB);
\draw[-,  draw=darkgreen\opp] (GenerationSourceC) to (NodeEC);
\draw[-,  draw=darkgreen\opp] (GenerationSourceC) to ($(NodeBB)!0.5!(NodeEC)$);

\begin{scope}
   \coordinate[below left=1.4cm and 0cm of c1] (IN);
   \coordinate[right=19.5cm of IN] (OUT);
   \draw[draw=cyan!20, line width=20pt, arrows = {-Stealth[inset=0pt, angle=30:60pt]}] (IN) to (OUT);

    \node [text=black, font=\Large, anchor = west] at (IN) {Input Data};

    \node [text=black, font=\Large] at ($(OUT)-(2.8,0)$) {Knowledge Graph};
    
    \node [text=black, font=\Large] (D2N) at ($(IN)!0.45!(OUT)$) {Data2Neo};

\end{scope}

    \node [anchor=west] at  ($(midc)-(0.73,2.1)$) (LegendEntry1) { \small User Defined Component:};
    \node [ anchor=south east, fill=orange!10,minimum width=8mm, minimum height=4mm, right=0.1cm of LegendEntry1.east] {};
\end{tikzpicture}

%% file: assets/tikz/example_conversion.tex
\newcommand{\internodedist}{1.8cm}
\newcommand{\intertabledist}{0.5cm}
\newcommand{\rellabel}[2]{{\footnotesize #1} \\ {\footnotesize #2}}
\newcommand{\kgnode}[7]{
    \ifstrequal{#2}{below}{
        \node [fill=#6, label={[align=left,label distance=-2mm]#2:\textcolor{#6!40!black}{\textbf{\normalsize #1}} \\ \normalsize #3 },#4] (#1) #5 {};
    }{
        \node [fill=#6, label={[label distance=-2mm, align=left#7]#2:\textcolor{#6!40!black}{\textbf{\normalsize #1}} \\[1mm] \normalsize #3 },#4] (#1) #5 {};
    }
}

\begin{tikzpicture}
\begin{scope}
    \footnotesize
    \tikzset{
        table/.style={
            font=\footnotesize,
            anchor=west,
            matrix of nodes,
             nodes={
                draw,
                align=left, 
                minimum height=0.4cm,
                minimum width=1.8cm,
                anchor=west, 
             },
            row sep=-\pgflinewidth,
            column sep=-\pgflinewidth,
        },
        label/.style={
            font=\bfseries,
            draw,
            align=left,
            minimum width=1.8cm,
            fill=gray!20, 
        },
        arrow/.style={->, dashed, shorten >=1pt, shorten <=1pt},
    }

    \pgfdeclarehorizontalshading{c2e}{100bp}{
        color(0bp)=(c!20);
        color(50bp)=(e!20)
    }

    \pgfdeclarehorizontalshading{b2c}{100bp}{
        color(0bp)=(b!20);
        color(50bp)=(c!20)
    }

    \matrix (ProductsTable) [table, nodes={minimum width=2.1cm}] {
        |[label, shading=c2e, minimum width=2.1cm]| \normalsize Product \\ 
        ID \\
        Name \\
        CategoryCode \\
        SupplierID \\
    };

    \matrix (SupplierTable) [table, right=\intertabledist of ProductsTable] {
        |[label, fill=d!20]|\normalsize  Supplier \\ 
        ID \\
        Name \\
    };

    \matrix (OrdersTable) [table, below=0cm of ProductsTable] {
        |[label, fill=b!20]|\normalsize  Order \\ 
        ID \\
        Date \\
        EmployeeID \\
    };
    
    \matrix (EmployeesTable) [table, right=\intertabledist of OrdersTable, nodes={minimum width=2.0cm}] {
        |[label,  minimum width=2.0cm, fill=a!20]| \normalsize Employee \\ 
        ID \\
        Name \\
        BirthDate \\
    };

    \matrix (OrdersDetailTable) [table, below=0cm of OrdersTable, nodes={minimum width=2.2cm}] {
        |[label, minimum width=2.2cm, shading=b2c]| \normalsize OrderDetail \\ 
        OrderID \\
        ProductID \\
        Amount \\
    };

    \draw[arrow] (ProductsTable-5-1.east) -- (SupplierTable-2-1.west);
    \draw[arrow] (OrdersTable-4-1.east) -- (EmployeesTable-2-1.west);
    \draw[arrow] (OrdersDetailTable-2-1.west) to[out=-250,in=250] (OrdersTable-2-1.west);
    \draw[arrow] (OrdersDetailTable-3-1.west) to[out=-250,in=250] (ProductsTable-2-1.west);

\end{scope}

    \tikzstyle{bigarrow} = [single arrow, minimum width=2cm, minimum height=1cm, single arrow head extend=0.2cm, fill=gray!30]

    \node[bigarrow, above right=-5mm and 4mm of EmployeesTable] {};

\begin{scope}[
        every node/.style={circle,thick,minimum size=8mm,anchor=center}    
    ]

    \kgnode{Employee}{right}{$\cdot$ id \\ \normalsize  $\cdot$ name}{above right=-5mm and 15mm of SupplierTable}{}{a}{,yshift=-3mm}
    \kgnode{Order}{below}{ $\cdot$ id \\  \normalsize $\cdot$ date}{below=\internodedist of Employee}{}{b}{,yshift=-3mm}
    \kgnode{Supplier}{right}{ $\cdot$ id \\  \normalsize $\cdot$ name}{right=\internodedist of Employee}{}{d}{,yshift=-3mm}
    \kgnode{Product}{right}{ $\cdot$ name}{below=\internodedist of Supplier}{}{c}{,yshift=-2mm, xshift=1mm}
    \coordinate (midpointcity) at ($(Employee.east)!0.5!(Supplier.west)$);

    \coordinate (midpointcat) at ($(Order.east)!0.7!(Product.west)$);
    \kgnode{Category}{right}{$\cdot$ name \\  \normalsize $\cdot$ conversion\_date}{below=\internodedist*0.7 of midpointcat}{}{e}{,yshift=-4.5mm}

\end{scope}

\begin{scope}[
        every edge/.style={draw=black,thick}
    ]

    \path [->] (Employee) edge node[below, rotate=90, align=center] {\rellabel{SELLS}{}} (Order) {};
    \path [->] (Order) edge node[below, align=center] {\rellabel{CONTAINS}{\normalsize $\cdot$ amount}} (Product) {};
    \path [->] (Supplier) edge node[below,rotate=90, align=center] {\rellabel{SUPPLIES}{}} (Product) {};
    \path [->] (Product) edge node[below, sloped, align=center] {\rellabel{IN}{}} (Category) {};

\end{scope}

\end{tikzpicture}